\documentclass[twocolumn,aps,prx,floatfix,superscriptaddress]{revtex4-1}

\usepackage{graphicx}
\usepackage{dcolumn}
\usepackage{epsfig}
\usepackage{amssymb}
\usepackage{amsmath}
\usepackage{natbib}
\usepackage{array}
\usepackage{bbold}
\usepackage{color}

\begin{document}

\title{Zeeman spectroscopy of excitons and hybridization of electronic states \\ in few-layer WSe$_2$, MoSe$_2$ and MoTe$_2$}

\author{Ashish Arora}
\email{arora@uni-muenster.de}
\affiliation{Laboratoire National des Champs Magn\'{e}tiques Intenses, CNRS-UGA-UPS-INSA-EMFL, 25 rue des Martyrs, 38042 Grenoble, France}
\affiliation{Institute of Physics and Center for Nanotechnology, University of M\"unster, Wilhelm-Klemm-Strasse 10, 48149 M\"unster, Germany}
\author{Maciej Koperski}
\affiliation{Laboratoire National des Champs Magn\'{e}tiques Intenses, CNRS-UGA-UPS-INSA-EMFL, 25 rue des Martyrs, 38042 Grenoble, France}
\affiliation{School of Physics and Astronomy, University of Manchester, Oxford Road, Manchester, M13 9PL, United Kingdom}
\affiliation{National Graphene Institute, University of Manchester, Oxford Road, Manchester, M13 9PL, United Kingdom}
\author{Artur Slobodeniuk}
\affiliation{Laboratoire National des Champs Magn\'{e}tiques Intenses, CNRS-UGA-UPS-INSA-EMFL, 25 rue des Martyrs, 38042 Grenoble, France}
\author{Karol~Nogajewski}
\affiliation{Laboratoire National des Champs Magn\'{e}tiques Intenses, CNRS-UGA-UPS-INSA-EMFL, 25 rue des Martyrs, 38042 Grenoble, France}
\affiliation{Institute of Experimental Physics, Faculty of Physics, University of Warsaw, Pasteura 5, 02-093, Warszawa, Poland}
\author{Robert Schmidt}
\affiliation{Institute of Physics and Center for Nanotechnology, University of M\"unster, Wilhelm-Klemm-Strasse 10, 48149 M\"unster, Germany}
\author{Robert~Schneider}
\affiliation{Institute of Physics and Center for Nanotechnology, University of M\"unster, Wilhelm-Klemm-Strasse 10, 48149 M\"unster, Germany}
\author{Maciej R. Molas}
\affiliation{Laboratoire National des Champs Magn\'{e}tiques Intenses, CNRS-UGA-UPS-INSA-EMFL, 25 rue des Martyrs, 38042 Grenoble, France}
\affiliation{Institute of Experimental Physics, Faculty of Physics, University of Warsaw, Pasteura 5, 02-093, Warszawa, Poland}
\author{Steffen~Michaelis~de~Vasconcellos}
\affiliation{Institute of Physics and Center for Nanotechnology, University of M\"unster, Wilhelm-Klemm-Strasse 10, 48149 M\"unster, Germany}
\author{Rudolf Bratschitsch}
\affiliation{Institute of Physics and Center for Nanotechnology, University of M\"unster, Wilhelm-Klemm-Strasse 10, 48149 M\"unster, Germany}
\author{Marek Potemski}
\affiliation{Laboratoire National des Champs Magn\'{e}tiques Intenses, CNRS-UGA-UPS-INSA-EMFL, 25 rue des Martyrs, 38042 Grenoble, France}
\affiliation{Institute of Experimental Physics, Faculty of Physics, University of Warsaw, Pasteura 5, 02-093, Warszawa, Poland}

\begin{abstract}
Monolayers and multilayers of semiconducting transition metal dichalcogenides (TMDCs) offer an ideal platform to explore valley-selective physics with promising applications in valleytronics and information processing. Here we manipulate the energetic degeneracy of the $\mathrm{K}^+$ and $\mathrm{K}^-$ valleys in few-layer TMDCs. We perform high-field magneto-reflectance spectroscopy on WSe$_2$, MoSe$_2$, and MoTe$_2$ crystals of thickness from monolayer to the bulk limit under magnetic fields up to 30~T applied perpendicular to the sample plane. Because of a strong spin-layer locking, the ground state A excitons exhibit a monolayer-like valley Zeeman splitting with a negative $g$-factor, whose magnitude increases monotonically when thinning the crystal down from bulk to a monolayer.  Using the $\mathbf{k\cdot p}$ calculation, we demonstrate that the observed evolution of $g$-factors for different materials is well accounted for by hybridization of electronic states in the $\mathrm{K}^+$ and $\mathrm{K}^-$ valleys. The mixing of the valence and conduction band states induced by the interlayer interaction decreases the $g$-factor magnitude with an increasing layer number. The effect is the largest for MoTe$_2$, followed by MoSe$_2$, and smallest for WSe$_2$.

\end{abstract}

\maketitle

Hybridization of electronic states in van der Waals-coupled layers of semiconducting transition metal dichalcogenides (TMDCs), significantly affects their energy bands and optical properties. Most striking is a dramatic change in the quasiparticle band gap character, from a direct bandgap at the $\mathrm{K}$-point of the Brillouin zone in monolayers to an indirect $\Gamma-\Lambda$ band gap in multilayers and bulk crystals \cite{1,2}. In contrast, the energy of the optical band gap, which is due to $\mathrm{K}$-point excitons in any mono-, multi- and bulk-crystals rather weakly depends on the number of layers in TMDC stacks \cite{1}. This effect is due to both the hybridization of electronic states at the $\mathrm{K}$-points \cite{3} and the change in the dielectric environment with different number of layers \cite{4}. While the hybridization of the electronic states leads to (often unresolved) multiplets of intralayer (electron and hole within the same layer) and spatially-separated interlayer excitons (electron and hole confined to different layers), the dielectric environment largely determines the excitonic binding energy and the optical band gap. The hybridization of electronic states in TMDC multilayers is also encoded in the magnitudes of the effective Land\'{e} $g$-factors of the coupled states. However, in contrast to the energetic positions of electronic resonances, $g$-factors are less sensitive to the effects of Coulomb interaction (dielectric environment) \cite{5}.

In TMDC monolayers, the band structure at the $\mathrm{K}$-point consists of energetically degenerate states at the $\mathrm{K}^+$ and $\mathrm{K}^-$ valleys. However, the two valleys possess opposite magnetic moments, and can be individually addressed using $\sigma^+$ and $\sigma^-$-polarized light \cite{1}. An externally applied magnetic field in the Faraday geometry lifts the valley degeneracy, resulting in a so-called valley Zeeman splitting \cite{1}. Therefore, the $g$-factors of the excitons can be measured using helicity-resolved spectroscopy under magnetic fields \cite{6,7,8,9,10,11,12,13,14,15,16,17,18,19}. In multilayer and bulk TMDCs, it has been found that the spin orientation of the carriers is strongly coupled to the valleys within the individual layers (``spin-layer locking'') \cite{17,19,20,21}. Therefore, many salient features of monolayer physics are preserved in multilayers. As a consequence, intralayer excitons form with their characteristic negative $g$-factors \cite{17,21}. Moreover, spin-layer locking effects have recently enabled the unambiguous identification of interlayer excitons in bulk TMDCs with positive $g$-factors \cite{17,19}. However, a systematic investigation of the effect of layer number and the hybridization of electronic states on the valley Zeeman effect has not been reported so far.

Here, we perform circular polarization-resolved micro-reflectance contrast ($\mu$RC) spectroscopy on 2H-WSe$_2$, 2H-MoSe$_2$ and 2H-MoTe$_2$ crystals of variable thickness (from monolayer to bulk) under high magnetic fields of up to $B=30$ T and at a temperature of $T=4$ K. We measure the layer thickness-dependent valley Zeeman splittings of the ground state A excitons ($X_A^{1s}$) and compare the observed trends with the $\mathbf{k\cdot p}$ theory. The model takes into account the interlayer admixture of valence and conduction bands and corrections from the higher and lower bands from adjacent layers at the $\mathrm{K}$-point of the Brillouin zone. We find that the hybridization of the electronic states at the band extrema has profound effects on the $g$-factors of the excitons. Overall, the exciton $g$-factor decreases with an increasing layer thickness where the extent of this reduction depends upon the magnitude of interlayer interaction in the TMDCs.

\section{Experiment}

Monolayer and few-layer flakes of TMDCs are mechanically exfoliated \cite{22} onto SiO$_2$(80nm)/Si substrates. The layer number in the MoSe$_2$ and WSe$_2$ crystals is determined by the optical contrast, Raman spectroscopy and the low-temperature (liquid helium) micro-photoluminescence \cite{23,24,25}. For MoTe$_2$, the thickness characterization was performed using ultra-low frequency Raman spectroscopy \cite{26,27,28}, in addition to the reflectance contrast and atomic force microscopy (AFM) measurements (see Fig.~\ref{MOKEvsB} in Appendix~\ref{Sec:experiment}).

Magneto-reflectance measurements are performed using a fiber-based low-temperature probe inserted inside a resistive magnet with 50 mm bore diameter, where magnetic fields up to 30 T are generated in the center of the magnet. Light from a tungsten halogen lamp is routed inside the cryostat using an optical fiber of 50 $\mu$m diameter and focused on the sample to a spot of about 10 $\mu$m diameter with an aspheric lens of focal length 3.1 mm (numerical aperture NA=0.68). The sample is displaced by $x-y-z$ nano-positioners. The reflected light from the sample is circularly polarized using the combination of a quarter wave plate (QWP) and a polarizer. The emitted polarized light is collected using an optical fiber of 200 $\mu $m diameter, dispersed with a monochromator and detected using a liquid nitrogen cooled Si CCD (WSe$_2$ and MoSe$_2$) or InGaAs array (MoTe$_2$). During the measurements, the configuration of QWP-polarizer assembly is kept fixed, producing one state of circular polarization, whereas the effect corresponding to the other polarization state can be measured by reversing the direction of magnetic field, as a result of the time reversal symmetry \cite{11,29}.

We define the reflectance contrast $C(\lambda)$ at a given wavelength $\lambda$ as $C(\lambda)=[R(\lambda)-R_0 (\lambda)]/[R(\lambda)+R_0(\lambda)]$, where $R_0(\lambda)$ is the reflectance spectrum of the SiO$_2$/Si substrate and $R(\lambda)$ is the one of the TMDC flake kept on the substrate. $C(\lambda)$ spectral line shapes are modeled using a transfer matrix method-based approach to obtain the transition energies \cite{30}. The excitonic contribution to the dielectric response function is assumed to follow a Lorentz oscillator-like mode \cite{5,31}
\begin{equation}
\epsilon(E)=(n_b+ik_b)^2+\sum_{j}\frac{A_j}{E_{0j}^2-E^2-i\gamma_jE},
\end{equation}
where $n_b+ik_b$ is the background complex refractive index of the TMDC being investigated, which excludes excitonic effects, and is kept equal to that of bulk material (WSe$_2$ \cite{32}, MoSe$_2$ \cite{33}, or MoTe$_2$ \cite{33} in the respective cases). $E_0$, $A$ and $\gamma$ are the transition energy, the oscillator strength parameter, and the full width at half maximum (FWHM) linewidth parameter, whereas the index $j$ represents the sum over excitons.
\begin{figure}[t!]
\includegraphics[width=8 cm]{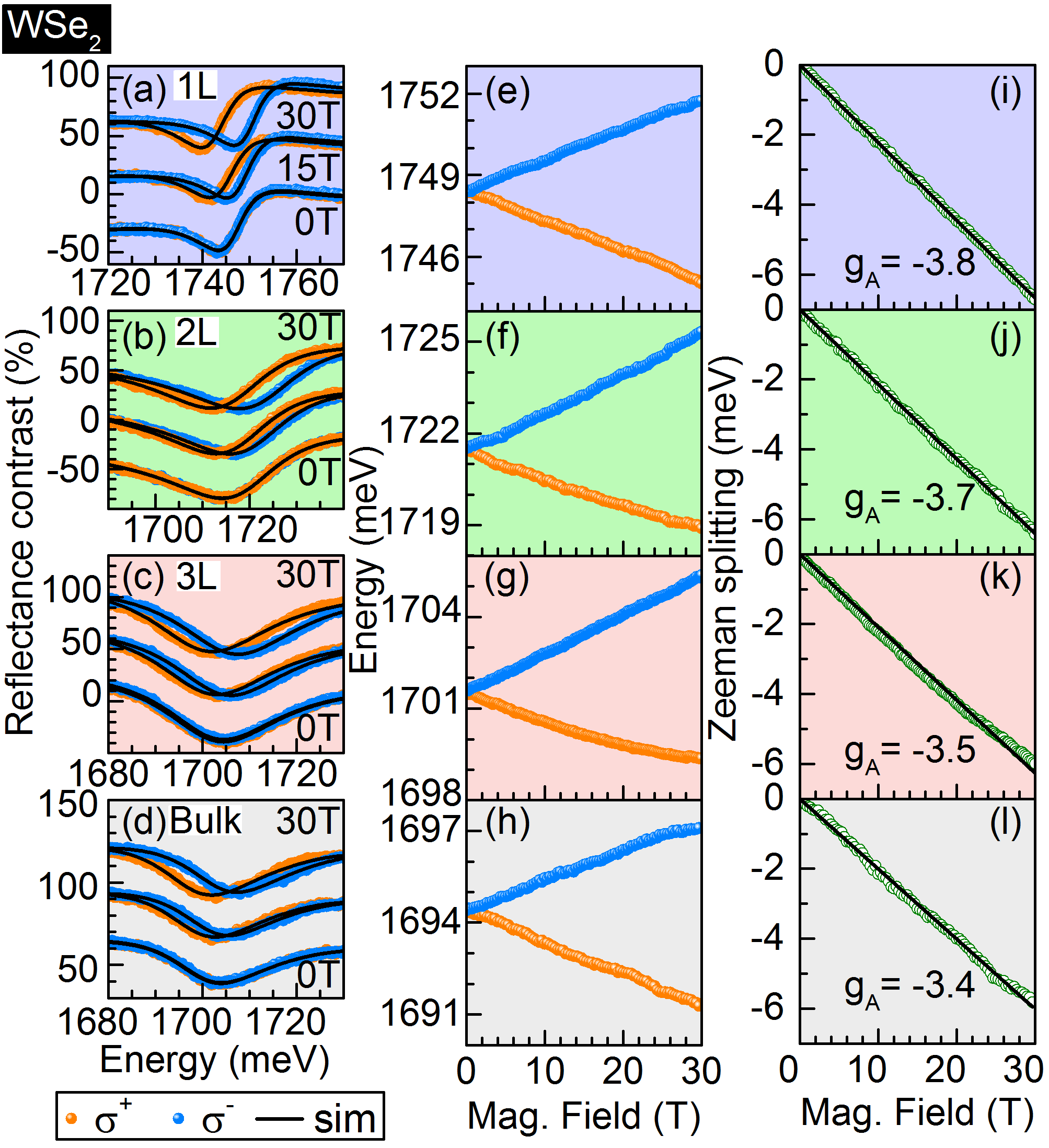}
\centering
\caption{(a)-(d) Helicity-resolved microreflectance contrast spectra of the ground state A excitons ($X_A^{1s}$) in 1L, 2L, 3L and bulklike WSe$_2$ crystals, respectively, measured at a temperature of T = 4.2 K under magnetic fields of 0 T, 15 T, and 30 T. Orange and blue spheres represent the experimental data for the $\sigma^+$ and $\sigma^-$ polarizations, respectively, whereas solid lines are the modeled spectra. The curves for $B>0$ T are shifted vertically with respect to the $B=0$ T measurement for clarity. (e)-(h) Excitonic transition energies derived for the two circular polarizations from the modeled spectra in (a)-(d), respectively, as a function of magnetic field from 0 to 30 T. (i)-(l) Green circles represent the Zeeman splittings for the corresponding cases in (e)-(h), respectively, whereas solid lines are linear fits to the data.}
\label{figure1}
\end{figure}

Figures 1(a)-(d) depict $\sigma^+$ (orange) and $\sigma^-$ (blue) components of the $\mu$RC spectra of the ground state A exciton ($X_A^{1s}$) in 1L, 2L, 3L and bulklike WSe$_2$ crystals kept at liquid He temperature of 4.2 K under magnetic fields of 0 T, 15 T and 30 T. With increasing magnetic field, one clearly observes an energetic splitting between the two circular components of excitonic features, indicating the Zeeman effect. The spectra are modeled (solid black lines) using the transfer matrix method as described before. The derived $X_A^0$ transition energies for the two circular polarizations for the four cases are displayed in Figs. 1(e)-(h). The excitonic Zeeman splittings are defined as
$\Delta E_X=E_{\sigma^+}-E_{\sigma^-}=g_X\mu_B B$, where $E_{\sigma^+}$ and $E_{\sigma^-}$ are the transition energies for the two circular polarizations, $g_X$ is the exciton’s effective $g$-factor and $\mu_B$ is the Bohr's magneton (0.05788 meV/T). The Zeeman splittings calculated from Figs. 1(e)-(h) are shown in Figs. 1(i)-(l) respectively (green circles), as a function of magnetic field.
Fig. 2 displays the corresponding data for the 1L to 3L thick MoSe$_2$ while Fig.~3 shows the plots for 2L, 3L and 4L MoTe2. The magnitude of the splitting increases linearly with rising magnetic field in all cases.

The excitonic $g$-factors obtained from the above analysis are summarized in Table~1 and plotted in Fig.~4.
The 1L and 40 nm thick bulklike MoTe$_2$ crystals, whose $g$-factors ($-4.8\pm0.2$ \cite{11} and $-2.4\pm0.1$ \cite{17} respectively) are also marked in Fig.~4, are obtained from the same single crystalline source material on SiO$_2$/Si substrates, as the 2L, 3L, and 4L samples. For bulk MoSe$_2$, the $g$-factor was measured for a 30 nm thick flake exfoliated on sapphire substrate \cite{19}. Interestingly, the absolute value of the $g$-factor for $X_A^0$ clearly decreases monotonically with increasing layer thickness and approaches the limiting bulk value of $-3.4\pm0.1$, $-2.7\pm0.1$, and $-2.4\pm0.1$ for WSe$_2$, MoSe$_2$ \cite{19}, and MoTe$_2$ \cite{17}, respectively.
\begin{figure}[t!]
\includegraphics[width=8 cm]{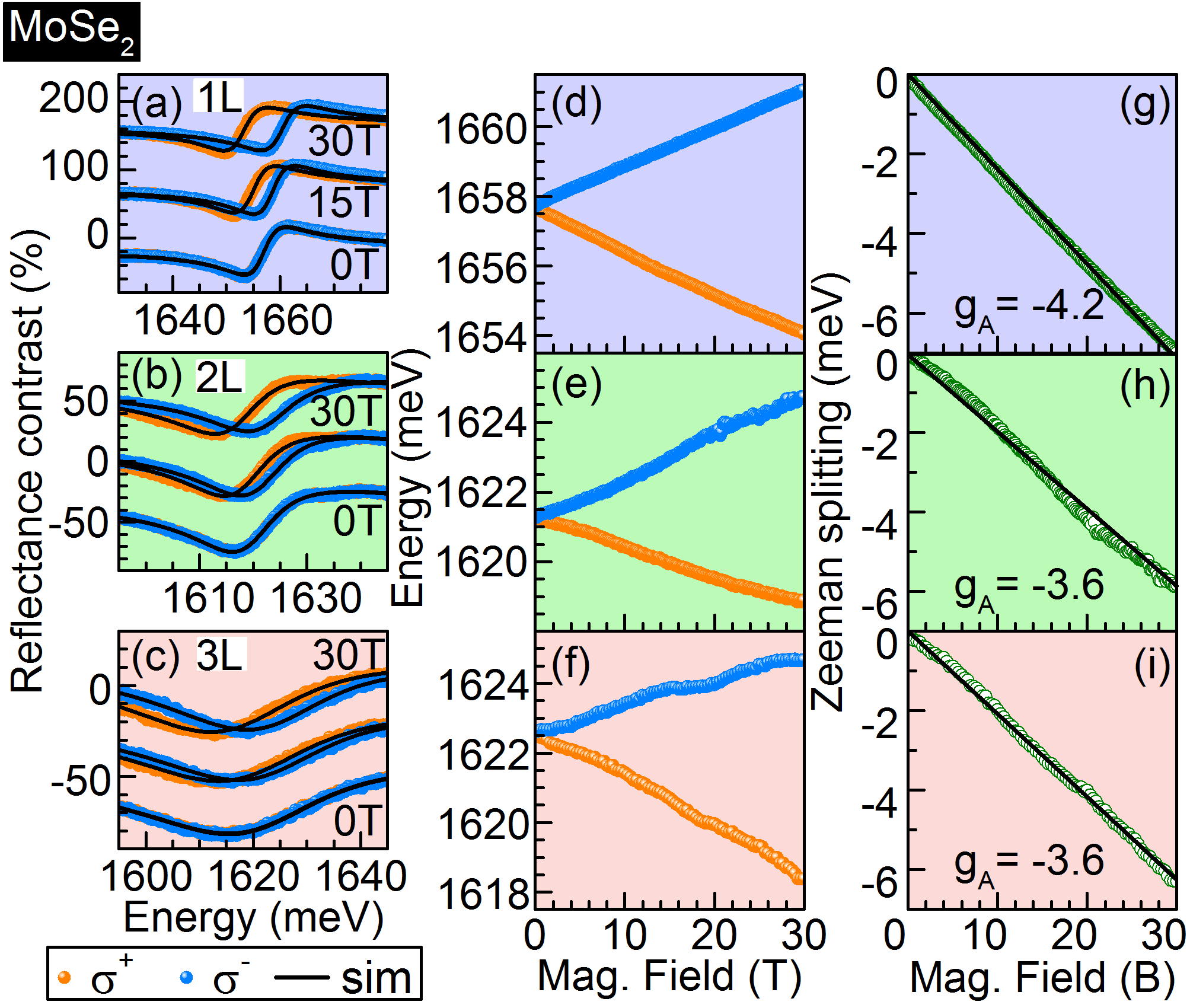}
\centering
\caption{(a)-(c) Helicity-resolved microreflectance contrast spectra of the ground state A excitons ($X_A^{1s}$) in 1L, 2L and 3L MoSe$_2$ crystals, respectively, at a temperature of
T = 4.2~K under magnetic fields of 0 T, 15 T, and 30 T. Orange and blue spheres represent the experimental data for the $\sigma^+$ and $\sigma^-$ polarizations respectively, whereas solid lines are the modeled spectra. The curves for $B>0$ T are shifted vertically with respect to the $B=0$ T measurement for clarity. (d)-(f) Excitonic transition energies derived for the two circular polarizations from the modeled spectra in (a)-(c), respectively, as a function of magnetic field from 0 to 30 T. (g)-(i) Green circles represent the Zeeman splittings for the corresponding cases in (d)-(f), respectively, whereas solid lines are linear fits to the data.}
\label{figure2}
\end{figure}
\begin{figure}[t!]
\includegraphics[width=8 cm]{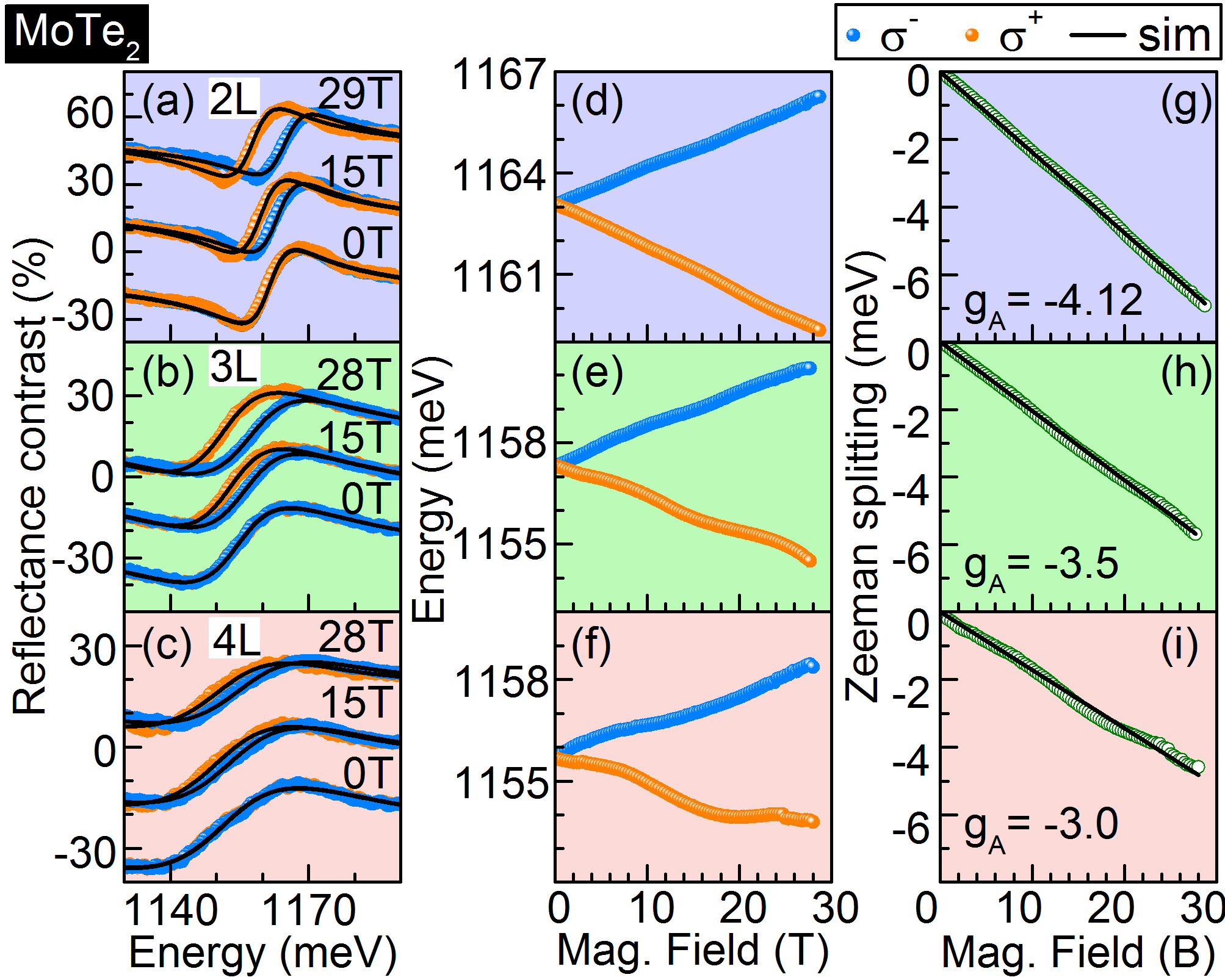}
\centering
\caption{(a)-(c) Helicity-resolved microreflectance contrast spectra of the ground state A excitons ($X_A^{1s}$) in 2L, 3L and 4L MoTe$_2$ crystals, respectively, at a temperature of
 T = 4.2 K under different magnetic fields. Orange and blue spheres represent the experimental data for the $\sigma^+$ and $\sigma^-$ polarizations respectively, whereas solid lines are the modeled spectra. The curves for $B>0$ T are shifted vertically with respect to the $B=0$ T measurement for clarity. (d)-(f) The excitonic transition energies obtained from modeling the reflectance contrast spectra in (a)-(c), respectively. (g)-(i) Green circles represent the Zeeman splittings for the corresponding cases in (d)-(f), respectively, whereas solid lines are linear fits to the data.}
\label{figure3}
\end{figure}
\begin{table}[h]
 \begin{center}
 \begin{tabular}{p{1.5cm}  c p{5mm} c  p{5mm} c}
 \hline\hline \\ [-0.5ex]
 {Layer thickness}  &   WSe$_2$ & &	MoSe$_2$ &&	MoTe$_2$   \\ [1.0ex]
 \hline \\
 1L    &  $-3.8\pm0.1$ & &	$-4.2\pm0.1$	& & $-4.8\pm0.2$ \\ [1.5ex]
 2L    &  $-3.7\pm0.1$ & &	$-3.6\pm0.1$	& & $-4.12\pm0.05$        \\ [1.5ex]
 3L    &  $-3.5\pm0.1$ & &	$-3.6\pm0.1$	& & $-3.5\pm0.1$          \\ [1.5ex]
 4L    &       -       & &        -         & & $-3.0\pm0.1$	       \\ [1.5ex]
 Bulk  &  $-3.4\pm0.1$ & &	$-2.7\pm0.1$    & & $-2.4\pm0.1$ \\ [1.5ex]
 \hline
\end{tabular}
\end{center}
\caption{\label{tab:gfactors}
Effective $g$-factors of the ground state A exciton $X_A^{1s}$ measured in WSe$_2$, MoSe$_2$ and MoTe$_2$ crystals of variable thickness at a temperature of T=4.2 K.}
\end{table}

\section{Theory}

The experimental results presented above demonstrate a significant deviation of $A$-exciton $g$-factor in the multilayer TMDC from the one found in monolayer. Moreover, absolute value of the $g$-factor decreases monotonically with the number of layers $N$. Although weak, the hybridization of $\mathrm{K}$-electronic states is likely at the origin of this effect. In order to confirm this hypothesis we develop a $\mathbf{k\cdot p}$ theory \cite{13,34,35,36} both for mono- and multilayers and derive exciton $g$-factors in this framework. Namely we consider the properties of the quasiparticles in the corners of the 1-st Brillouin zone (where the studied optical transitions take place). Next, we focus on $\mathrm{K}^+$ point for brevity.

Let us first consider the TMDC monolayer, situated in $xy$ plane. Electronic excitations in $\mathrm{K}^+$ point of such a system are described by a set of Bloch states $\{|\Psi_n,s\rangle\}$ with energies $\{E_{ns}\}$. The subscript $n$ enumerates the bands, and index $s=\uparrow,\downarrow$ determines their spin degrees of freedom.
According to $\mathbf{k\cdot p}$ method, the quasiparticles with the momentum $\mathbf{k}=(k_x,k_y)$ near $\mathrm{K}^+$ point are described  by the matrix elements $\langle\Psi_n,s|\widehat{H}^{(1)}|\Psi_{n'},s'\rangle$ of one-particle Hamiltonian
\begin{align}
\widehat{H}^{(1)}(\boldsymbol{\rho},z)=& \frac{\mathbf{\widehat{p}}^2}{2m_0}+ U(\boldsymbol{\rho},z) + \nonumber \\ +&
\frac{\hbar}{4m_0^2c^2}\big[\nabla U(\boldsymbol{\rho},z),\mathbf{\widehat{p}}\big]\boldsymbol{\sigma} +
\frac{\hbar}{m_0}\mathbf{k\widehat{p}}.
\end{align}
Here $m_0$ is electron's mass, $c$ --- speed of light, $\hbar$ --- Planck's constant and
$\boldsymbol{\sigma}=(\sigma_x, \sigma_y, \sigma_z)$ are Pauli matrices.
We also introduced in-plane coordinate $\boldsymbol{\rho}=(x,y)$, the momentum operator $\mathbf{\widehat{p}}=-i\hbar\nabla$
and the crystal field of a monolayer $U(\boldsymbol{\rho},z)$.
The first two terms of the Hamiltonian define the energies $E_n$ of the bands, doubly degenerated by spin.
The next part describes the spin-orbital interaction.
It lifts the spin-degeneracy of $n$-th band by the value $\Delta_n$ ({\it i.e.} in total, one has $E_n\pm \Delta_n/2$ for $s=\uparrow,\downarrow$ states respectively).
The last $\mathbf{k\widehat{p}}$ term couples different Bloch states of monolayer.
The coupling gives rise to additional energy of the $n$-th band $\delta E_n = g_n\mu_B B$ in the
presence of magnetic field $\mathbf{B}=B\mathbf{e}_z$.
Here $\mu_B$ is the Bohr magneton. According to the Roth formula \cite{37} the
spin-independent $g$-factor of the $n$-th band is
\begin{equation}
g_n\!=\!\frac{1}{2m_0}\!\!\sum_{n'\neq n}\frac{|\langle\Psi_n,s|\widehat{p}_+|\Psi_{n'},s\rangle|^2-
|\langle\Psi_n,s|\widehat{p}_-|\Psi_{n'},s\rangle|^2}{E_n-E_{n'}},
\end{equation}
where $\widehat{p}_\pm=\widehat{p}_x\pm i\widehat{p}_y$. The interaction of electron's
magnetic moment with magnetic field gives the spin correction $\delta E_s=\sigma_s\mu_BB$,
where $\sigma_s=+1(-1)$ for $s=\uparrow(\downarrow)$.
Finally the energy of the $n$-th band in $\mathrm{K}^+$ point is
$E_{ns}(B)=E_n+\sigma_s\Delta_n/2 + g_n\mu_BB + \sigma_s\mu_BB$.
In this picture, the experimentally measured $A$-exciton $g$-factor is doubled difference
$g_{exc}=2(g_c-g_v)$ between the $g$-factors of conduction $g_c$ and valence $g_v$ bands.
We take this result as a reference point for our next calculations.

The $N$-layer TMDC crystal with $2\text{H}$ stacking order can be represented as a pile of
monolayers separated by a distance $l$. Each successive layer of such a crystal is $180^\circ$ rotated with
respect to the previous one. The one-particle Hamiltonian for this system has a form
\begin{align}
\widehat{H}^{(N)}&(\mathbf{\boldsymbol{\rho}},z)=\frac{\mathbf{\widehat{p}}^2}{2m_0}+ \sum_{m=1}^{N}U\big((-1)^{m+1}\mathbf{\boldsymbol{\rho}},z-z_m\big)
+ \nonumber \\ +& \sum_{m=1}^{N}\frac{\hbar}{4m_0^2c^2}
\big[\nabla U\big((-1)^{m+1}\boldsymbol{\rho},z-z_m\big),\mathbf{\widehat{p}}\big]\boldsymbol{\sigma}
+\frac{\hbar}{m_0}\mathbf{k\widehat{p}}
\end{align}
It contains a sum of potentials from all the layers, with coordinates $z_m=(m-1)l$.
The potential of each even stratum has a form $U(-\boldsymbol{\rho},z-z_{2m})$.
The sign ``-'' before two-dimensional coordinate $\boldsymbol{\rho}$ represents the fact
of $180^\circ$ rotation. Note that the orientation of the first layer of the system
does not depend on $N$. Hence, it is convenient to match the $\mathrm{K}^+$ point of
any multilayer with $\mathrm{K}^+$ point of its lowest part. This uniquely determines
the form of the unperturbed Bloch states in each $m$-th stratum of the system.
We consider the set of such states in $\mathrm{K}^+$ point as a new basis
$\{|\Psi_n^{(m)},s\rangle\}$ of the multilayer.
The states $|\Psi_n^{(1)},s\rangle$ belong to the lowest (first) stratum and
are equal to $|\Psi_n,s\rangle$ by definition.
The other part of the basis can be derived from $|\Psi_n^{(1)},s\rangle$ with the help
of crystal symmetry operations (see Appendix~\ref{Sec:theory}).
We suppose the orthogonality of states from different layers $\langle\Psi_n^{(m)},s|\Psi_{n'}^{(m')},s'\rangle=\delta_{nn'}\delta_{mm'}\delta_{ss'}$.

According to our choice of the basis, the bands of multilayer are $N$-times degenerated (in leading approximation). The Roth formula is not applicable in this case. To solve this problem, we apply the L\"{o}wdin partitioning technique \cite{37} to multimatrix $\langle\Psi_n^{(m)},s|\widehat{H}^{(N)}|\Psi_{n'}^{(m')},s'\rangle$ and derive the effective conduction $H^{(N)}_{cs}$ and valence $H^{(N)}_{vs}$ band Hamiltonians. They act in the spaces, spanned over $\{|\Psi_c^{(m)},s\rangle\}$ and $\{|\Psi_v^{(m)},s\rangle\}$ basis states respectively and can be presented as $N\times N$ matrices. Their eigenvalues determine the multilayer $g$-factors.

The effective Hamiltonians have a form of pentadiagonal matrix.
Their main diagonal contains $E_n\pm\Delta_n/2$ terms, spin term $\delta E_s$ and $B$-dependent correction
$\delta E_n^{(m)}=g^{(m)}_n\mu_BB$ to the energies of quasiparticles from $m$-th layer.
Here and further in the text $n$ takes $c$ or $v$ values.
The corresponding $g$-factor of $m$-th layer originates from $\mathbf{k\widehat{p}}$ term and reads
\begin{align}
g^{(m)}_n=&\frac{1}{2m_0}\sum_{n'\neq n}\sum_{\eta=\pm}\frac{\eta|\langle\Psi^{(m)}_n,s|\widehat{p}_\eta
|\Psi^{(m)}_{n'},s\rangle|^2}{E_n-E_{n'}} + \nonumber \\
+&\frac{1}{2m_0}\sum_{\langle\!\langle m',m\rangle\!\rangle}\sum_{n'\neq n}\sum_{\eta=\pm}
\frac{\eta|\langle\Psi^{(m)}_n,s|\widehat{p}_\eta|\Psi^{(m')}_{n'},s\rangle|^2}{E_n-E_{n'}}.
\end{align}
The result is a sum of intralayer and interlayer contributions. The intralayer part is nothing but the
monolayer's $g$-factor considered above. The interlayer part determines the deviation $\propto\delta g_n$
from this $g$-factor. The symbol $\langle\!\langle m',m\rangle\!\rangle$ describes the nearest neighbours
of the $m$-th stratum. Namely $\langle\!\langle m',1\rangle\!\rangle \rightarrow m'=2$;
$\langle\!\langle m',N\rangle\!\rangle\rightarrow m'=N-1$ and
$\langle\!\langle m',m\rangle\!\rangle\rightarrow m'=m-1, m+1$ for $m=2,3,\dots N-1$.
We restrict our summation in such a way, since the next nearest neighbour terms are
suppressed by the distance between the layers. We omit these terms from our study.

The sub- and superdiagonal matrix elements of considered Hamiltonians describe the admixing of the
Bloch states between neighbour layers. For conduction bands they origin from $\mathbf{k\widehat{p}}$ terms and
have a linear in $\propto k_x\pm ik_y$ dependence. For valence bands they appear from the crystal
field of neighbour layers and are proportional to material dependent constant $t\sim 40\dots 70\,\text{meV}$ \cite{38}.
The next nearest sub- and superdiagonal matrix elements are linear in magnetic field $\propto \bar{g}_n\mu_BB$
and also originate from $\mathbf{k\widehat{p}}$ terms.

In our model, we did several simplifications:
{\it i}) Only the intralayer matrix elements of spin-orbit interaction are taken into account in $\widehat{H}^{(N)}(\mathbf{\boldsymbol{\rho}},z)$.
The interlayer terms are beyond of accuracy of our approximation;
{\it ii}) The interlayer crystal field corrections to the bands energy positions are supposed to be small
and omitted from our study;
{\it iii}) The $\mathbf{k}$ dependent part of spin-orbital interaction
$\propto\big[\nabla U(\boldsymbol{\rho},z),\mathbf{k}\big]\boldsymbol{\sigma}$ is neglected. This term produces a small correction to the spin
$g$-factor, which is beyond the scope of this paper;
{\it iv}) The effective Hamiltonians are considered up to the linear in $\mathbf{k}$ terms.
The higher order corrections give zero contribution to the band energy in $\mathrm{K}^+$ point,
and therefore are not important in this study.

The diagonalization of $H^{(N)}_{cs}$ and $H^{(N)}_{vs}$ matrices provides a new set of energy bands with
corresponding eigenstates. Hence, we expect a series of exciton lines instead of single $A$-exciton one.
It is well known that the exciton lines in the optical spectra of TMDCs as-exfoliated on substrates such as SiO$_2$ or sapphire
have a significantly large inhomogeneous line width broadening compared to the homogeneous line widths \cite{39,40,41}.
In principle, it is possible to achieve the homogeneous linewidth with hBN, which might able to resolve the close-lying
individual lines of excitons in multilayers \cite{39,40,41}. However, in the present case, we calculate
the average $g$-factor from all lines and compare it with the experiment. The corresponding observable
(see the detailed $\mathbf{k\cdot p}$ analysis for each multilayer in Appendix~\ref{Sec:theory}) as a function of number of layers $N$ has the following form
\begin{equation}
g^{(N)}_{exc}= g_{exc} + 4\Big(1-\frac{1}{N}\Big)\big[\delta g_v -\delta g_c - g_u\big] + O\Big(\frac{t^2}{\Delta_v^2}\Big).
\end{equation}
The parameters $\delta g_c$ and $\delta g_v$ are the interlayer corrections to the conduction and valence band energies, $g_u$ and $t$ appear from the interlayer admixture of the conduction and valence band states respectively. This formula indicates the measured dependence of the exciton $g$-factor, if we suppose
$\delta g_v-\delta g_c-g_u>0$.
\begin{figure}[t!]
\includegraphics[width=5.5 cm]{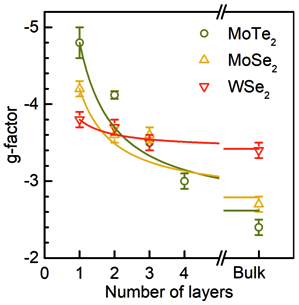}
\centering
\caption{Effective $g$-factors of the ground state A excitons ($X_A^{1s}$) in WSe$_2$, MoSe$_2$, and MoTe$_2$ as a function of layer thickness from monolayer to the bulk limit. The $g$-factors for 1L and bulk MoTe$_2$ are taken from Refs.~\cite{11} and \cite{19}, respectively, and were measured on a flake obtained from the same crystal as used in the present work, and under the same experimental conditions. Solid lines represent the theoretical model as described in the main text.}
\label{figure4}
\end{figure}

Using Eq.~6, we fit the experimental data in Fig.~4 to the first order (solid lines). Here, we fix $g_{exc}$ to the experimentally measured $g$-factor of the monolayer. The deviation of the fits from the experimental data could be explained by the neglected second-order term $O(t^2/\Delta_v^2)$. Apart from this, our model predicts the correct qualitative trends of the $g$-factors observed in the experiment as a function of layer thickness. The fitting parameter $[\delta g_v-\delta g_c-g_u]$ is found to be equal to 0.1, 0.35 and 0.55 for WSe$_2$, MoSe$_2$ and MoTe$_2$, respectively. A larger value of this parameter in MoSe$_2$ and MoTe$_2$ points towards a stronger interlayer interaction in these materials, when compared to that of WSe$_2$. This is in agreement with the {\it ab-initio} calculations where the spin-valley coupling of holes to a particular layer was found to be significantly larger than (comparable to) the interlayer hopping in W-based (Mo-based) compounds \cite{38}. Furthermore, an increased interlayer coupling has been reported when the chalcogen atom changes from Se to Te \cite{42}. The recent observation of spatially indirect (``interlayer'') excitons in bulklike MoTe$_2$ \cite{17} and MoSe$_2$ \cite{19}, where a large interlayer interaction results in a significant oscillator strength of interlayer excitons \cite{3,43} support our conclusions as well. Indeed, we find that the strength of interlayer excitons is much smaller in W-based TMDCs, which leads to an absence of their signature in the optical spectra of WS$_2$ and WSe$_2$ \cite{19}.

In summary, we have measured the Zeeman effect of intralayer A excitons in semiconducting WSe$_2$, MoSe$_2$, and MoTe$_2$ crystals of variable thickness from monolayer to the bulk limit, using helicity-resolved magneto-reflectance contrast spectroscopy under high magnetic fields up to 30 T. The magnitude of the negative $g$-factors of the A excitons displays a monotonic decrease as the layer thickness is increased from monolayer to a bulklike crystal. The effect is qualitatively explained with a model considering thickness-dependent interlayer interactions, and band mixing effects. Our results represent the first report devoted to the effect of the band hybridization on magneto-optics of multilayer TMDCs, and will contribute towards a better understanding of TMDCs along with future device-based applications.

\section{Acknowledgements}

The authors acknowledge the financial support from Alexander von Humboldt foundation, German
Research Foundation (DFG project no. AR 1128/1-1), European Research Council (MOMB project
no. 320590), the EC Graphene Flagship project (no. 604391) and the ATOMOPTO project
(TEAM programme of the Foundation for Polish Science cofinanced
by the EU within the ERDFund).

The authors declare no competing financial interest.

\appendix

\begin{widetext}

\section{Characterization of MoTe$_2$ crystals of different thickness}
\label{Sec:experiment}

Supplementary Fig. \ref{MOKEvsB}(a) shows Raman spectra of MoTe$_2$ crystals with thicknesses ranging from $1$L to $6$L and  $40$~nm think bulklike material. The Raman-active modes A$_{1g}$, E$_{2g}^1$, B$_{2g}^1$ as well as low-frequency shear modes are clearly visible \cite{27,28}. For initial characterization, $\mu$RC measurements on the flakes with layer thickness $1$L to $4$L and the bulklike crystal are performed at low temperatures in the absence of magnetic field (\textbf{B}=$0$). Fig. \ref{MOKEvsB}(b) displays the $\mu$RC spectra as a function of layer thickness. Features corresponding to the neutral (ground state $1$s) A exciton resonance $\rm X_A^{1s}$ and a broad B exciton resonance $\rm X_B^{1s}$ are identified in the spectra. A weak shoulder at the high-energy side of $\rm X_A^{1s}$ is associated with the excited state $\rm X_A^{2s}$ exciton resonance. An additional feature at $1.183$ eV for the bulklike flake arises due to the optically active interlayer $\rm X_{IL}$ exciton \cite{17}. The derived excitonic transition energies for the various observed features are shown in Fig. \ref{MOKEvsB}(c). The $\rm X_A^{1s}$ resonance undergoes a red shift from $1.196 \pm 0.001$ eV to $1.1307 \pm 0.001$ eV as the layer thickness is increased from $1$L to bulk as has been observed previously also in WSe$_2$ \cite{25}, MoSe$_2$ \cite{23}, WS$_2$ \cite{3} and MoTe$_2$ \cite{26,Ruppert2014}. At the same time, the energy difference between the $\rm X_A^{1s}$ and $\rm X_A^{2s}$ resonances decreases from $127$ meV to $25$ meV as the layer thickness is increased from $2$L to bulk ($\rm X_A^{2s}$ is not observed for the $1$L flake). This behavior points towards a reduction of the excitonic binding energy with increasing crystal thickness. It has been largely associated with an increasing dielectric constant when the layer thickness is increased \cite{Cheiw2012}, and has also been observed previously in WSe$_2$ \cite{25} and MoSe$_2$ \cite{23}. It is worth mentioning that the binding energies of the A excitons in $1$L and bulklike MoTe$_2$ have been calculated to be $710$ meV \cite{Ramasu2012} and $150$ meV \cite{17}, respectively.
\begin{figure*}[t!]
\includegraphics[width=13 cm]{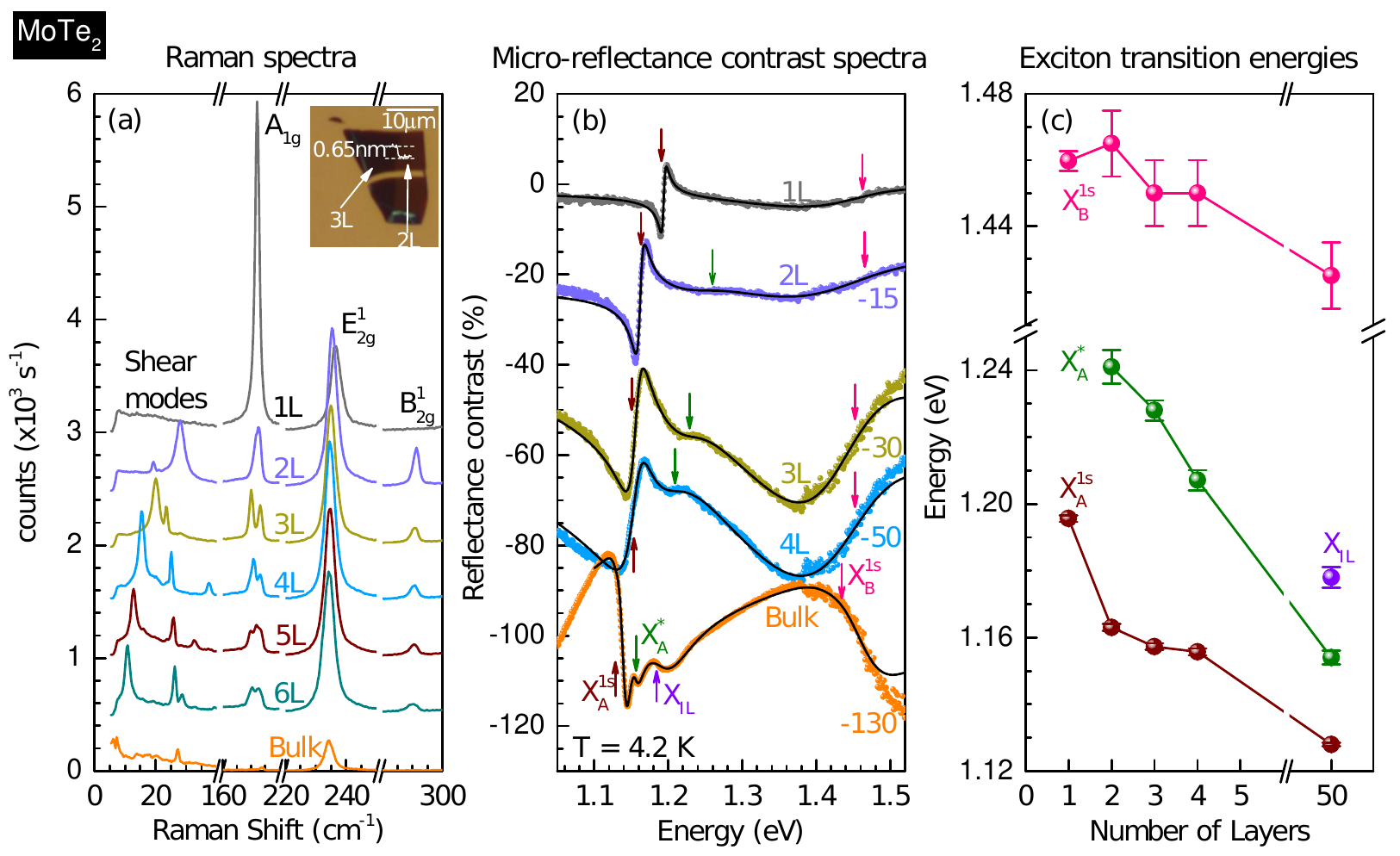}
\centering
\caption{(a) Raman spectra for MoTe$_2$ crystals with thickness ranging from monolayer (1L) to bulklike (40 nm). The typical Raman-active modes $\rm A_{1g}$, $\rm E_{2g}^1$, $\rm B_{2g}^1$, as well as low-frequency shear modes are observed. The inset shows the optical microscope image of a flake consisting of $2$L and $3$L thick areas, along with a line profile measured by atomic force microscopy used for the height determination of the monolayer. (b) Micro-reflectance contrast ($\mu$RC) spectra obtained for MoTe$_2$ as a function of layer thickness with the ground state ($\rm X_A^{1s}$ and $\rm X_B^{1s}$) and first excited state $\rm X_A^{2s}$ exciton transition. For bulk MoTe$_2$, the interlayer exciton resonance $\rm X_{IL}$ is observed as well. (c) Transition energies of the measured exciton resonances as a function of layer thickness, derived by modeling the spectra as explained in the text.} \label{MOKEvsB}
\end{figure*}

\section{Theory}
\label{Sec:theory}

Our purpose is to calculate the $g$-factors of $A$-excitons in $\mathrm{K}^\pm$
valleys of multilayer TMDC. In order to do it, we extend the 7-band
$\mathbf{k\cdot p}$ model \cite{34,35,13} to $N$-layer case,
derive the effective Hamiltonians and then calculate the positions of energy bands
as a function of magnetic field. Namely, we use the monolayer Bloch functions to construct the basis states of a multilayer.
Then, we compute first-order $\mathbf{k\cdot p}$ and spin-orbital corrections to the Hamiltonian of the system.
Finally, we derive the effective valence and conduction bands Hamiltonians
as a function of external magnetic field, diagonalize them and find the bands $g$-factors.
In further, we consider $\mathrm{K}^+$ point for brevity.

\subsection{Monolayer}

The 7-band model contains 3 additional bands below the valence band and 2 bands above the conduction one \cite{34,35,13}.
The Bloch states in $\mathrm{K}^+$ point of monolayer are $|\Psi_{v-3},s\rangle, \,\, |\Psi_{v-2},s\rangle, \,\, |\Psi_{v-1},s\rangle,\,\, |\Psi_v,s\rangle,\,\, |\Psi_c,s\rangle,\,\,
|\Psi_{c+1},s\rangle, \,\, |\Psi_{c+2},s\rangle$.
The lower index $n=v-3, v-2,\dots c+1$ indicates the band, $s=\uparrow,\downarrow$ is the spin degree of freedom.
The basis vectors are defined as a decomposition $|\Psi_n,s\rangle=|\Psi_n\rangle|s\rangle$.
They can be classified according to irreducible representations of the symmetry group of the crystal \cite{34,35}. All the group transformations are based on the in-plane $2\pi/3$ rotation $C_3$  and in-plane mirror
reflection $\sigma_h$. The states $|\Psi_{v-3},s\rangle, \,\, |\Psi_v,s\rangle, \,\, |\Psi_c,s\rangle,
\,\, |\Psi_{c+2},s\rangle$ are even under mirror transformation, while the $|\Psi_{v-2},s\rangle, \,\, |\Psi_{v-1},s\rangle, \,\, |\Psi_{c+1},s\rangle$ are odd. The $\mathbf{k\cdot p}$ perturbation terms couple only the states with the same parity. Therefore the odd states do not affect $g$-factors of monolayer and can be excluded from this particular case.
Taking into account the transformation properties of the remaining states \cite{13}
$C_3|\Psi_v,s\rangle=|\Psi_v,s\rangle$, $C_3|\Psi_c,s\rangle=\omega^*|\Psi_c,s\rangle$, $C_3|\Psi_{v-3},s\rangle=\omega|\Psi_{v-3},s\rangle$, $C_3|\Psi_{c+2},s\rangle=\omega|\Psi_{c+2},s\rangle$
with $\omega=e^{2i\pi/3}$, one obtains $\mathbf{k\cdot p}$ matrix elements, presented in Table~\ref{tab:monolayer} \cite{34,35,13,Kormanyos2014}.
 \begin{table*}[h]
 \begin{center}
 \begin{tabular}{p{1.5cm}  p{1.5cm}  p{1.5cm}  p{1.5cm}  p{1.2cm}}
 \hline\hline \\ [-0.5ex]
 {$H_{\mathbf{kp}}$} & $|\Psi_v,s\rangle$ & $|\Psi_c,s\rangle$  & $|\Psi_{v-3},s\rangle$ & $|\Psi_{c+2},s\rangle$  \\ [1.0ex]
 \hline \\
 $|\Psi_v,s\rangle$  & $E_v$  & $\gamma_3k_+$ & $\gamma_2k_-$ & $\gamma_4k_-$  \\ [1.5ex]
 $|\Psi_c,s\rangle$  & $\gamma^*_3k_-$ & $E_c$ & $\gamma_5k_+$ & $\gamma_6k_+$ \\ [1.5ex]
 $|\Psi_{v-3},s\rangle$  & $\gamma^*_2k_+$ & $\gamma^*_5k_-$ & $E_{v-3}$ & 0 \\ [1.5ex]
 $|\Psi_{c+2},s\rangle$  & $\gamma^*_4k_+$ & $\gamma^*_6k_-$ & 0 & $E_{c+2}$ \\ [1.5ex]
 \hline
\end{tabular}
\end{center}
\caption{\label{tab:monolayer}
Non-zero $\mathbf{k\cdot p}$ matrix elements of monolayer. }
\end{table*}
Here we introduced notation $k_\pm=k_x\pm ik_y$ and a set of energies $\{E_n\}$ in
$\mathrm{K}^+$ point for clarity.
The spin-orbit interaction, considered as a perturbation, gives the correction
$\sigma_s\Delta_n/2$ to diagonal elements of the table, with $\sigma_s=+1(-1)$ for
$\uparrow(\downarrow)$ states.
Applying the L\"{o}wdin procedure we calculate the energies of $c$ and $v$
bands in $\mathrm{K}^+$ point
\begin{equation}
E_{cs}(B)=E_c+\sigma_s\Delta_c/2+g_c\mu_B B + \sigma_s\mu_BB,
\quad E_{vs}(B)=E_v+\sigma_s\Delta_v/2+g_v\mu_B B + \sigma_s\mu_BB.
\end{equation}
Here $\mu_B$ is the Bohr magneton, $B$ is the strength of magnetic field $\mathbf{B}=B\mathbf{e}_z$ and
\begin{eqnarray}
g_v&=&\frac{2m_0}{\hbar^2}\left[\frac{|\gamma_3|^2}{E_c-E_v}+\frac{|\gamma_2|^2}{E_v-E_{v-3}}
+\frac{|\gamma_4|^2}{E_v-E_{c+2}}\right], \\
g_c&=&\frac{2m_0}{\hbar^2}\left[\frac{|\gamma_3|^2}{E_c-E_v}-\frac{|\gamma_5|^2}{E_c-E_{v-3}}
-\frac{|\gamma_6|^2}{E_c-E_{c+2}}\right].
\end{eqnarray}
The last term in $E_{cs}(B)$ and $E_{vs}(B)$ is a free electron Zeeman energy.
In our study, we suppose the spin-orbital corrections to electron's magnetic moment are small \cite{1}.

Note that the $A$-exciton transitions in $\mathrm{K}^+$ points are possible only in $\sigma^+$
circularly polarized light. In magnetic field their energy shifts by the value $\delta E_+=(g_c-g_v)\mu_B B$.
In $\mathrm{K}^-$ point, transitions are active only in $\sigma^-$ polarization and are characterised by the shift $\delta E_-=-\delta E_+$, which is a consequence of time reversal symmetry in the system.
Therefore the measurable exciton $g$-factor is
\begin{eqnarray}
g_{exc}=2(g_c-g_v)=-\frac{4m_0}{\hbar^2}\left[\frac{|\gamma_5|^2}{E_c-E_{v-3}}
+\frac{|\gamma_6|^2}{E_c-E_{c+2}}+\frac{|\gamma_2|^2}{E_v-E_{v-3}}
+\frac{|\gamma_4|^2}{E_v-E_{c+2}}\right].
\end{eqnarray}
We use this result as a reference point for our next calculations.

\subsection{Bilayer}

A bilayer TMDC crystal with 2$H$ stacking order can be presented as two monolayers
separated by distance $l$, with the second (upper) layer $180^\circ$ rotated relative to the
first (lower) one. It is convenient to arrange them in $z=-l/2$ and $z=l/2$ planes respectively.
In this presentation the crystal has the inverse symmetry $I$ with
the inversion center placed in the middle between the monolayers.

There are two subsets of basis Bloch states in $\mathrm{K}^+$ point of bilayer --
from the lower and upper strata.
The first part $\{|\Psi^{(1)}_n,s\rangle\}$ coincides with the Bloch states of monolayer
$\{|\Psi_n,s\rangle\}$, located in $z=-l/2$ plane.
The second part $\{|\Psi^{(2)}_n,s\rangle\}$ can be derived as
$|\Psi^{(2)}_n,s\rangle=p_nK_0I|\Psi^{(1)}_n,s\rangle$. Here $K_0$ is the
conjugation operator and $p_n=\pm1$ is the parity of $|\Psi^{(1)}_n\rangle$.
As a result the upper states are transformed as a complex conjugated
to the lower ones. It leads to opposite optical selection rules
for such states. Namely, in $\mathrm{K}^+$ point of bilayer the first (second) layer
absorbs only $\sigma^+(\sigma^-)$ polarized light respectively. Hence, the bilayer does not
possess any optical dichroism, which is nothing but a manifestation of the inversion
symmetry of the crystal.

In contrast to the monolayer case, the odd states of bilayer give non-zero $\mathbf{k\cdot p}$ contributions.
Therefore, taking into account their rotational $C_3|\Psi_{v-2},s\rangle=\omega^*|\Psi_{v-2},s\rangle$, $C_3|\Psi_{v-1},s\rangle=\omega|\Psi_{v-1},s\rangle$, $C_3|\Psi_{c+1},s\rangle=\omega|\Psi_{c+1},s\rangle$
and inversion properties we derive the Table~\ref{tab:bilayer_even} and Table~\ref{tab:bilayer_odd}.

\begin{table}[h]
\begin{center}
 \begin{tabular}{p{1.5cm}  p{1.5cm}  p{1.5cm}  p{1.5cm} p{1.5cm}  p{1.5cm}  p{1.5cm}  p{1.5cm}  p{1.2cm}}
 \hline\hline \\ [-0.5ex]
 {$H_{\mathbf{kp}}$} & $|\Psi^{(1)}_v,s\rangle$ & $|\Psi^{(2)}_v,s\rangle$ & $|\Psi^{(1)}_c,s\rangle$  & $|\Psi^{(2)}_c,s\rangle$ & $|\Psi^{(1)}_{v-3},s\rangle$ & $|\Psi^{(1)}_{c+2},s\rangle$ & $|\Psi^{(2)}_{v-3},s\rangle$
 & $|\Psi^{(2)}_{c+2},s\rangle$  \\ [1.0ex]
 \hline \\ [0.1ex]

 $|\Psi^{(1)}_v,s\rangle$  & $E_v$ & $t$ & $\gamma_3k_+$  & $rk_-$ & $\gamma_2k_-$ &  $\gamma_4k_-$  & $ak_+$ & $bk_+$ \\ [1.5ex]

 $|\Psi^{(2)}_v,s\rangle$  & $t$ & $E_v$ & $rk_+$  & $\gamma_3k_-$ & $ak_-$ &  $bk_-$  & $\gamma_2k_+$ & $\gamma_4k_+$  \\ [1.5ex]

 $|\Psi^{(1)}_c,s\rangle$  & $\gamma^*_3k_-$ & $r^*k_-$ & $E_c$  & $uk_+$ & $\gamma_5k_+$ &  $\gamma_6k_+$  &
 0& 0 \\ [1.5ex]

 $|\Psi^{(2)}_c,s\rangle$  & $r^*k_+$ & $\gamma^*_3k_+$ & $uk_-$  & $E_c$  & 0 & 0&  $\gamma_5k_-$ & $\gamma_6k_-$  \\ [1.5ex]

 \hline
\end{tabular}
\end{center}
\caption{\label{tab:bilayer_even}
The $\mathbf{k\cdot p}$ matrix elements of bilayer between even states.}
\end{table}
Note that the diagonal matrix elements in the case of bi- and other multilayers should contain small corrections $\delta E_n$, which appear from the crystal field of adjacent layers. However, according to our rough estimation such diagonal terms produce less than 5\% deviation to the $g$-factors of multilayers, considered here. Therefore, for the clarity reasons we put $\delta E_n=0$ for this particular study, remembering, however, that these terms can give non-negligible corrections in other cases.
\begin{table}[h]
\begin{center}
 \begin{tabular}{p{1.5cm} p{1.5cm} p{1.5cm}  p{1.5cm}  p{1.5cm} p{1.5cm}  p{1.2cm}}
 \hline\hline \\ [-0.5ex]
 {$H_{\mathbf{kp}}$} & $|\Psi^{(1)}_{v-2},s\rangle$ & $|\Psi^{(1)}_{v-1},s\rangle$
 & $|\Psi^{(1)}_{c+1},s\rangle$ & $|\Psi^{(2)}_{v-2},s\rangle$
   & $|\Psi^{(2)}_{v-1},s\rangle$ & $|\Psi^{(2)}_{c+1},s\rangle$  \\ [1.0ex]
 \hline \\ [0.1ex]

 $|\Psi^{(1)}_v,s\rangle$  & 0 & 0 & 0 & $ck_-$ & $dk_+$ & 0\\ [1.5ex]

 $|\Psi^{(2)}_v,s\rangle$  & $-ck_+$ & $-dk_-$ & 0 & 0 & 0 & 0 \\ [1.5ex]

 $|\Psi^{(1)}_c,s\rangle$  & 0 & 0 & 0 & $fk_+$ & 0 & $jk_-$ \\ [1.5ex]

 $|\Psi^{(2)}_c,s\rangle$  & $-fk_-$ & 0 & $-jk_+$ & 0 & 0 & 0 \\ [1.5ex]
 \hline
\end{tabular}
\end{center}
\caption{\label{tab:bilayer_odd}
The $\mathbf{k\cdot p}$ matrix elements of bilayer between even and odd states.}
\end{table}
We also introduced the admixing parameter $t$ between valence bands of the first and second layers.
Then we add the spin orbit-interaction, apply the L\"{o}wding partitioning to corresponding matrix
elements and derive the effective valence and conduction band Hamiltonians.
The valence band Hamiltonian, written in the basis $\{|\Psi^{(1)}_v,s\rangle,|\Psi^{(2)}_v,s\rangle\}$, reads
\begin{equation}
H^{(2)}_{vs}=\left[
          \begin{array}{cc}
            E_v+\sigma_s\frac{\Delta_v}{2} & t  \\
            t & E_v-\sigma_s\frac{\Delta_v}{2}  \\
          \end{array}
        \right] +
        \left[
        \begin{array}{cc}
            g_v-\delta g_v +\sigma_s & 0 \\
            0 & -g_v+\delta g_v+\sigma_s \\
          \end{array}
        \right]\mu_BB.
\end{equation}
The conduction band Hamiltonian, written in the basis
$\{|\Psi^{(1)}_c,s\rangle,|\Psi^{(2)}_c,s\rangle\}$ is
\begin{equation}
H^{(2)}_{cs}=\left[
          \begin{array}{cc}
            E_c+\sigma_s\frac{\Delta_c}{2} & uk_+ \\
            uk_- & E_c-\sigma_s\frac{\Delta_c}{2}  \\
          \end{array}
        \right]+
        \left[
        \begin{array}{cc}
            g_c-\delta g_c +\sigma_s & 0  \\
            0 & -g_c+\delta g_c +\sigma_s \\
          \end{array}
        \right]\mu_BB.
\end{equation}
Note that spin-up and spin-down states can be considered separately.
The parameters $\delta g_v$ and $\delta g_c$ are the corrections to monolayer's $g$-factors of valence
and conduction bands
\begin{align}
\delta g_v=\frac{2m_0}{\hbar^2}\left[\frac{|a|^2}{E_v-E_{v-3}}-\frac{|b|^2}{E_{c+2}-E_v}
-\frac{|c|^2}{E_v-E_{v-2}}+\frac{|d|^2}{E_v-E_{v-1}}+\frac{|r|^2}{E_{c}-E_v}\right],
\end{align}
\begin{align}
\delta g_c=\frac{2m_0}{\hbar^2}\left[\frac{|f|^2}{E_c-E_{v-2}}+\frac{|j|^2}{E_{c+1}-E_c}
-\frac{|r|^2}{E_c-E_v}\right].
\end{align}
Technically, these corrections originate from additional non-zero $\mathbf{k\cdot p}$ matrix
elements between the states of bilayer, allowed by the symmetry.
The expressions for valence band energies up to $O(B)$ terms are
\begin{align}
E^\text{I}_{vs}=E_v+\sigma_s\sqrt{\frac{\Delta_v^2}{4}+t^2}+\frac{(g_v-\delta g_v)\Delta_v}{\sqrt{\Delta_v^2+4t^2}}\mu_B B+\sigma_s\mu_B B, \\
E^\text{II}_{vs}=E_v-\sigma_s\sqrt{\frac{\Delta_v^2}{4}+t^2}-\frac{(g_v-\delta g_v)\Delta_v}{\sqrt{\Delta_v^2+4t^2}}\mu_B B+\sigma_s\mu_B B.
\end{align}
The following eigenstates are
\begin{align}
&|\Phi^\text{I}_{vs}\rangle=\cos(\theta/2)|\Psi^{(1)}_v,s\rangle + \sigma_s\sin(\theta/2)|\Psi^{(2)}_v,s\rangle, \\
&|\Phi^\text{II}_{vs}\rangle=-\sigma_s\sin(\theta/2)|\Psi^{(1)}_v,s\rangle +\cos(\theta/2)|\Psi^{(2)}_v,s\rangle,
\end{align}
where we introduced $\{\cos\theta,\sin\theta\}=\{\Delta_v/\sqrt{\Delta_v^2+4t^2}, 2t/\sqrt{\Delta_v^2+4t^2}\}$.
The first state corresponds mostly to the optical transitions in $\sigma^+$ polarized light,
while the second one is active predominantly in $\sigma^-$ polarization. The intensity of emitted light
in $\mathrm{K}^+$ point is the same in both polarizations, which reflects the presence of inversion symmetry
of the bilayer crystal. The new conduction band energies are
\begin{align}
E^\text{I}_{cs}=E_c+\sigma_s\frac{\Delta_c}{2}-\sigma_sg_u\mu_BB+(g_c-\delta g_c)\mu_BB+\sigma_s\mu_BB,\\
E^\text{II}_{cs}=E_c-\sigma_s\frac{\Delta_c}{2}-\sigma_sg_u\mu_BB-(g_c-\delta g_c)\mu_BB+\sigma_s\mu_BB,
\end{align}
where $g_u=2m_0u^2/\hbar^2\Delta_c$.
The conduction band eigenstates with the same energies coincide with $|\Psi^{(1)}_{c},s\rangle$ and $|\Psi^{(2)}_{c},s\rangle$, up to $O(\mathbf{k}^2)$ order.
An analysis of new possible interband transitions demonstrates two $A$-exciton lines in $\mathrm{K}^+$ point of the bilayer. They are active in $\sigma^+$ and $\sigma^-$ polarisations respectively, and have opposite
energy shifts in magnetic field $\delta E_+=-\delta E_-=g^{(2)}_{exc}\mu_B B/2$. Here
\begin{equation}
g_{exc}^{(2)}=-2g_u+2(g_c-\delta g_c)
-2(g_v-\delta g_v)\frac{\Delta_v}{\sqrt{\Delta_v^2+4t^2}}
\end{equation}
is the $A$-exciton $g$-factor of the bilayer.
We rewrite this result up to $O(t^2/\Delta_v^2)$ order
\begin{equation}
g^{(2)}_{exc}=g_{exc} + 2\big[\delta g_v-\delta g_c-g_u\big] + \frac{4t^2}{\Delta_v^2}\big(g_v-\delta g_v\big),
\end{equation}
where $g_{exc}$ is the $A$-exciton $g$-factor of the monolayer.
The quantitative estimate of $\delta g_c$ and $\delta g_v$ deviations can be done in
numerical simulations and is beyond the scope of this study.  However, we will use
the experimental fact that for a bilayer $\delta g_v-\delta g_c-g_u>0$ and demonstrate
the self-consistency the other multilayer $g$-factors with this assumption.

\subsection{Trilayer}

We calculate the $g$-factors of a trilayer in the same way as in the bilayer. We introduce the three sets
of basis states $\{|\Psi^{(1)}_n\rangle\}$, $\{|\Psi^{(2)}_n\rangle\}$, $\{|\Psi^{(3)}_n\rangle\}$, which
belong to the layers $z=-l$, $z=0$ and $z=l$ respectively.
In this case the crystal has the mirror symmetry,
with the mirror plane $z=0$. This helps us to determine the following symmetry relations between the states
\begin{equation}
|\Psi^{(3)}_n\rangle=p_n\sigma_h|\Psi^{(1)}_n\rangle,\quad
|\Psi^{(2)}_n\rangle=p_n\sigma_h|\Psi^{(2)}_n\rangle.
\end{equation}
The states from $1$-st and $3$-d layer have the same rotational properties as in monolayer.
The rotational properties of the states from $2$-nd layer are complex conjugated to previous ones.
Therefore the $\mathbf{k\cdot p}$ matrix elements can be restored from the known result
\begin{align}
\langle\Psi^{(2)}_n|H_{\mathbf{kp}}|\Psi^{(3)}_m\rangle=
p_np_m\langle\Psi^{(2)}_n|H_{\mathbf{kp}}|\Psi^{(1)}_m\rangle, \\
\langle\Psi^{(3)}_n|H_{\mathbf{fkp}}|\Psi^{(3)}_m\rangle=
p_np_m\langle\Psi^{(1)}_n|H_{\mathbf{kp}}|\Psi^{(1)}_m\rangle.
\end{align}
We also assume $\langle\Psi^{(1)}_n|H_{\mathbf{kp}}|\Psi^{(3)}_m\rangle=0$
because of the large distance $2l$ between $1$-st and $3$-d layers.
Note that the $1$-st and $3$-d layers in $\mathrm{K}^+$ point absorb predominantly
$\sigma^+$ polarized light, while the $2$-nd layer absorbs $\sigma^-$ polarized light.
The Hamiltonian of trilayer TMDC can be written separately for spin-up and
spin-down states.
The Hamiltonian for valence bands, written in the basis
$\{|\Psi^{(1)}_v,s\rangle,|\Psi^{(2)}_v,s\rangle,|\Psi^{(3)}_v,s\rangle\}$, is
\begin{equation}
H^{(3)}_{vs}=\left[
          \begin{array}{ccc}
            E_v+\sigma_s\frac{\Delta_v}{2} & t & 0  \\
            t & E_v-\sigma_s\frac{\Delta_v}{2} & t  \\
            0 & t & E_v+\sigma_s\frac{\Delta_v}{2}  \\
          \end{array}
        \right] +
        \left[
        \begin{array}{ccc}
            g_v-\delta g_v+\sigma_s & 0  & \bar{g}_v\\
            0 & -g_v+2\delta g_v+\sigma_s & 0 \\
            \bar{g}_v & 0 & g_v-\delta g_v+\sigma_s \\
          \end{array}
        \right]\mu_B B.
\end{equation}
The Hamiltonian for conduction bands, written in the basis
$\{|\Psi^{(1)}_c,s\rangle,|\Psi^{(2)}_c,s\rangle,|\Psi^{(3)}_c,s\rangle\}$, reads
\begin{equation}
H^{(3)}_{cs}=\left[
          \begin{array}{ccc}
            E_c+\sigma_s\frac{\Delta_c}{2} & uk_+ & 0 \\
            uk_- & E_c-\sigma_s\frac{\Delta_c}{2} & uk_- \\
            0& uk_+&  E_c+\sigma_s\frac{\Delta_c}{2}\\
          \end{array}
        \right]+
        \left[
        \begin{array}{ccc}
            g_c-\delta g_c+\sigma_s & 0 & \bar{g}_c  \\
            0 & -g_c+2\delta g_c+\sigma_s & 0 \\
            \bar{g}_c & 0 & g_c-\delta g_c+\sigma_s  \\
          \end{array}
        \right]\mu_B B.
\end{equation}
Here we introduced
\begin{align}
\bar{g}_c=\frac{2m_0}{\hbar^2}\left[\frac{|f|^2}{E_c-E_{v-2}}+\frac{|j|^2}{E_{c+1}-E_c}
+\frac{|r|^2}{E_c-E_v}\right],
\end{align}
\begin{align}
\bar{g}_v=\frac{2m_0}{\hbar^2}\left[\frac{|r|^2}{E_v-E_c}-\frac{|a|^2}{E_v-E_{v-3}}
-\frac{|b|^2}{E_v-E_{c+2}}-\frac{|c|^2}{E_v-E_{v-2}}+\frac{|d|^2}{E_v-E_{v-1}}\right].
\end{align}
The new energy of conduction bands and corresponding eigenstates of such systems are
\begin{align*}
&E^{\text{I}}_{cs}=E_c-\sigma_s\frac{\Delta_c}{2}+
(-g_c+2\delta g_c+\sigma_s-2g_u\sigma_s)\mu_B B,
&|\Phi^\text{I}_{cs}\rangle&=|\Psi^{(2)}_{c},s\rangle; \\
&E^{\text{II}}_{cs}=E_c+\sigma_s\frac{\Delta_c}{2}+
(g_c-\delta g_c+\sigma_s-\bar{g}_c)\mu_B B, &|\Phi^\text{II}_{cs}\rangle&=\frac{1}{\sqrt{2}}\Big(|\Psi^{(1)}_{c},s\rangle-|\Psi^{(3)}_{c},s\rangle\Big);\\
&E^{\text{III}}_{cs}=E_c+\sigma_s\frac{\Delta_c}{2}+
(g_c-\delta g_c+\sigma_s-2g_u\sigma_s+\bar{g}_c)\mu_B B, &|\Phi^\text{III}_{cs}\rangle&=\frac{1}{\sqrt{2}}\Big(|\Psi^{(1)}_{c},s\rangle+|\Psi^{(3)}_{c},s\rangle\Big).
\end{align*}
The energies and normalised eigenstates of valence bands up to $O(t^2/\Delta_v^2)$ are
\begin{align*}
E^{\text{I}}_{vs}&=E_v-\sigma_s\frac{\Delta_v}{2}-2\frac{\sigma_st^2}{\Delta_v}+
(-g_v+2\delta g_v+\sigma_s)\mu_B B, &|\Phi^\text{I}_{vs}\rangle&=\Big(\frac{\sigma_st}{\Delta_v}|\Psi^{(1)}_{v},s\rangle-|\Psi^{(2)}_{v},s\rangle
+\frac{\sigma_st}{\Delta_v}|\Psi^{(1)}_{v},s\rangle\Big)\frac{\Delta_v}{\sqrt{\Delta_v^2+2t^2}}; \\
E^{\text{II}}_{vs}&=E_v+\sigma_s\frac{\Delta_v}{2}+
(g_v-\delta g_v+\sigma_s-\bar{g}_v)\mu_B B,  &|\Phi^\text{II}_{vs}\rangle&=\frac{1}{\sqrt{2}}\Big(|\Psi^{(1)}_{v},s\rangle-|\Psi^{(3)}_{v},s\rangle\Big);\\
E^{\text{III}}_{vs}&=E_v+\sigma_s\frac{\Delta_v}{2}+2\frac{\sigma_st^2}{\Delta_v}
+(g_v-\delta g_v+\sigma_s+\bar{g}_v)\mu_B B, &|\Phi^\text{III}_{vs}\rangle&=\Big(|\Psi^{(1)}_{v},s\rangle+2\frac{\sigma_st}{\Delta_v}|\Psi^{(2)}_{v},s\rangle+
|\Psi^{(1)}_{v},s\rangle\Big)\frac{\Delta_v}{\sqrt{2\Delta_v^2+4t^2}}.
\end{align*}

The lowest energy transitions in $\mathrm{K}_+$ point occur between new states with the same upper index.
The first transition is in $\sigma^-$ polarization, while the two others are in $\sigma^+$. The $g$-factors and normalised intensities of these transitions are
\begin{align*}
 &g^\text{I}=\frac{1}{\mu_B}\frac{d}{dB}(E^\text{I}_{c\downarrow}-E^\text{I}_{v\downarrow})=-(g_c-g_v+2\delta g_v -2\delta g_c-2g_u),  &J^\text{I}&=\frac{\Delta^2_v}{\Delta_v^2+2t^2};\\
 &g^\text{II}=\frac{1}{\mu_B}\frac{d}{dB}(E^\text{II}_{c\uparrow}-E^\text{II}_{v\uparrow})=g_c-g_v+\delta g_v -\delta g_c +\bar{g}_v-\bar{g}_c, &J^\text{II}&=1;\\
 &g^\text{III}=\frac{1}{\mu_B}\frac{d}{dB}(E^\text{III}_{c\uparrow}-E^\text{III}_{v\uparrow})=g_c-g_v+\delta g_v -\delta g_c-2g_u -\bar{g}_v+\bar{g}_c, &J^\text{III}&=\frac{\Delta_v^2}{\Delta_v^2+2t^2}.
\end{align*}
Therefore, since these three lines can not be resolved we introduce the average $A$-exciton $g$-factor
\begin{equation}
g^{(3)}_{exc}= g_{exc} + \frac{8}{3}\big[\delta g_v -\delta g_c - g_u\big] + \frac{4t^2}{9\Delta_v^2}\big(\delta g_c -\delta g_v + 4g_u + 3\bar{g}_v-3\bar{g}_c\big).
\end{equation}

\subsection{Quadrolayer}
We introduce four sets of Bloch states $\{|\Psi^{(1)}_n,s\rangle\}$,$\{|\Psi^{(2)}_n,s\rangle\}$,$\{|\Psi^{(3)}_n,s\rangle\}$,$\{|\Psi^{(4)}_n,s\rangle\}$ in $\mathrm{K}^+$ point of a quadrolayer. They correspond to $z=-3l/2$, $z=-l/2$, $z=l/2$ and $z=3l/2$ planes respectively. The quadrolayer possesses the inversion symmetry, which results into the following relations between the Bloch states
\begin{align}
|\Psi^{(4)}_n\rangle=p_nK_0I|\Psi^{(1)}_n\rangle,\quad |\Psi^{(3)}_n\rangle=p_nK_0I|\Psi^{(2)}_n\rangle.
\end{align}
Such relations allow to calculate $\mathbf{k\cdot p}$ matrix elements
\begin{align}
\langle\Psi^{(4)}_n|H_{\mathbf{kp}}|\Psi^{(3)}_m\rangle=
p_np_m\langle\Psi^{(1)}_n|H_{\mathbf{kp}}|\Psi^{(2)}_m\rangle, \\
\langle\Psi^{(4)}_n|H_{\mathbf{kp}}|\Psi^{(4)}_m\rangle=
p_np_m\langle\Psi^{(1)}_n|H_{\mathbf{kp}}|\Psi^{(1)}_m\rangle.
\end{align}
We also suppose that $\langle\Psi^{(4)}_n|H_{\mathbf{kp}}|\Psi^{(2)}_m\rangle=
\langle\Psi^{(4)}_n|H_{\mathbf{kp}}|\Psi^{(1)}_m\rangle=\langle\Psi^{(3)}_n|H_{\mathbf{kp}}|\Psi^{(1)}_m\rangle=0$ because the large
distance between the layers.
The Hamiltonian for valence bands, written in the basis
$\{|\Psi^{(1)}_v,s\rangle,|\Psi^{(2)}_v,s\rangle,|\Psi^{(3)}_v,s\rangle,|\Psi^{(4)}_v,s\rangle\}$,
can be presented as a sum of non-magnetic and magnetic parts $H^{(4)}_{vs}=\mathcal{H}^{(4)}_{vs}+ \mathcal{M}^{(4)}_{vs}\mu_B B$, with
\begin{equation}
\mathcal{H}^{(4)}_{vs}=\left[
          \begin{array}{cccc}
            E_v+\sigma_s\frac{\Delta_v}{2} & t & 0 & 0  \\
            t & E_v-\sigma_s\frac{\Delta_v}{2} & t & 0 \\
            0 & t & E_v+\sigma_s\frac{\Delta_v}{2} & t  \\
            0 & 0 & t & E_v-\sigma_s\frac{\Delta_v}{2}  \\
          \end{array}
        \right]
\end{equation}
and
\begin{equation}
\mathcal{M}^{(4)}_{vs}=
        \left[
        \begin{array}{cccc}
            g_v-\delta g_v+\sigma_s & 0 & \bar{g}_v & 0 \\
            0 & -g_v+2\delta g_v+\sigma_s & 0 & -\bar{g}_v \\
            \bar{g}_v & 0 & g_v-2\delta g_v+\sigma_s & 0 \\
            0 & - \bar{g}_v & 0 & -g_v+\delta g_v+\sigma_s \\
          \end{array}
        \right].
\end{equation}
The Hamiltonian for conduction bands, written in the basis
$\{|\Psi^{(1)}_c,s\rangle,|\Psi^{(2)}_c,s\rangle,|\Psi^{(3)}_c,s\rangle, |\Psi^{(4)}_c,s\rangle\}$,
has also the structure $H^{(4)}_{cs}=\mathcal{H}^{(4)}_{cs}+\mathcal{M}^{(4)}_{cs}\mu_B B$.
The corresponding matrices are
\begin{equation}
\mathcal{H}^{(4)}_{cs}=\left[
          \begin{array}{cccc}
            E_c+\sigma_s\frac{\Delta_c}{2} & uk_+ & 0 & 0\\
            uk_- & E_c-\sigma_s\frac{\Delta_c}{2} & uk_- & 0 \\
            0 & uk_+&  E_c+\sigma_s\frac{\Delta_c}{2} & uk_+  \\
            0& 0 & uk_-&  E_c-\sigma_s\frac{\Delta_c}{2}
          \end{array}
        \right],
\end{equation}
\begin{equation}
\mathcal{M}^{(4)}_{cs}=
        \left[
        \begin{array}{cccc}
            g_c-\delta g_c+\sigma_s & 0 & \bar{g}_c & 0 \\
            0 & -g_c+2\delta g_c+\sigma_s & 0 & -\bar{g}_c  \\
            \bar{g}_c & 0 & g_c-2\delta g_c+\sigma_s & 0 \\
            0& -\bar{g}_c & 0 & g_c-\delta g_c+\sigma_s  \\
          \end{array}
        \right].
\end{equation}
Note that, the valence and conduction band Hamiltonians for $N>4$ multilayers have the same
pentadiagonal strucure of their matrices. No one additional parameters appears for larger
TMDC crystals. The $A$-exciton $g$-factor is derived in analogues way as it is done for bi- and trilayer.
Since the expressions for eigenvalues and eigenstates are quite lengthy we present only the final result
\begin{equation}
g^{(4)}_{exc}=g_{exc}+3\big[\delta g_v -\delta g_c -g_u\big]+ O\Big(\frac{t^2}{\Delta_v^2}\Big).
\end{equation}

\subsection{Bulk}
The effective Hamiltonian of the bulk can be constructed in the same way as in bi-, tri- and quadrolayer.
Like in previous case, the spin-up and spin-down states can be considered separately. The effective Hamiltonian for valence band written in infinite basis
$\{\dots|\Psi^{(j-1)}_v,s\rangle,|\Psi^{(j)}_v,s\rangle,|\Psi^{(j+1)}_v,s\rangle\dots\}$ has the
matrix elements
\begin{align}
&\Big[H^{(\infty)}_{vs}\Big]_{j,j}=E_v+(-1)^{j+1}\Big[\sigma_s\frac{\Delta_v}{2}+(g_v-2\delta g_v)\mu_BB\Big] +
\sigma_s\mu_BB, \\
&\Big[H^{(\infty)}_{vs}\Big]_{j,j+1}=\Big[H^{(\infty)}_{vs}\Big]_{j+1,j}=t, \\
&\Big[H^{(\infty)}_{vs}\Big]_{j,j+2}=\Big[H^{(\infty)}_{vs}\Big]_{j+2,j}=(-1)^{j+1}\bar{g}_v\mu_B B.
\end{align}
The Hamiltonian for conduction band written in the basis
$\{\dots |\Psi^{(j-1)}_c,s\rangle,|\Psi^{(j+1)}_c,s\rangle,|\Psi^{(j+1)}_c,s\rangle,\dots\}$
has the matrix elements
\begin{align}
&\Big[H^{(\infty)}_{cs}\Big]_{j,j}=E_c+(-1)^{j+1}\Big[\sigma_s\frac{\Delta_c}{2} +
(g_c-2\delta g_c)\mu_BB\Big]+\sigma_s\mu_BB, \\
&\Big[H^{(\infty)}_{cs}\Big]_{2j\pm1,2j}=uk_+, \quad \Big[H^{(\infty)}_{cs}\Big]_{2j,2j\pm1}=uk_-, \\
&\Big[H^{(\infty)}_{cs}\Big]_{j,j+2}=\Big[H^{(\infty)}_{cs}\Big]_{j+2,j}=(-1)^{j+1}\bar{g}_c\mu_B B.
\end{align}
We solve the eigenvalues problem for a bulk in the following way. Let us consider a finite size
$N=2M$ multilayer with periodic boundary conditions. In this case, all the eigenstates of the crystal
can be parameterised by a wave-vector $k_n=\pi n/Ml$. Hereafter, we omit subscript $n$ for brevity and write $k$ instead of $k_n$.

We are looking for the valence band solutions in the form
\begin{equation}
|\Phi_{vs}^{k}\rangle=\frac{1}{\sqrt{M}}\sum_{m=1}^M e^{2ikml}\big[A_{vs}(k)|\Psi^{(2m-1)}_{v},s\rangle+
B_{vs}(k)|\Psi^{(2m)}_{v},s\rangle\big].
\end{equation}
This ansatz reduces the eigenvalues problem to
\begin{align}
&E(k)A_{vs}(k)=\Big\{E_v+\sigma_s\frac{\Delta_v}{2}+\big[g_v-2\delta g_v +\sigma_s+2\bar{g}_v\cos(2kl)\big]\mu_BB\Big\}A_{vs}(k) + 2te^{-ikl}\cos(kl)B_{vs}(k),\\
&E(k)B_{vs}(k)=\Big\{E_v-\sigma_s\frac{\Delta_v}{2} -\big[g_v-2\delta g_v -\sigma_s+2\bar{g}_v\cos(2kl)\big]\mu_BB\Big\}B_{vs}(k) +2te^{ikl}\cos(kl)A_{vs}(k).
\end{align}
The spectrum of the system up to $O(B)$ order is
\begin{equation}
E_{vs}^{\pm}(k)=E_v+\sigma_s\mu_BB\pm\frac12\sqrt{\Delta_v^2+16t^2\cos^2(kl)}\pm\sigma_s
\frac{\Delta_v[g_v-2\delta g_v+2\bar{g}_v\cos(2kl)]}{\sqrt{\Delta_v^2+16t^2\cos^2(kl)}}\mu_BB.
\end{equation}
Since we are interested in $A$-exciton transitions, we consider only the high energy bands.
The corresponding eigenstates up to zeroth order in magnetic field have the form
\begin{equation}
\left[
  \begin{array}{c}
    A^+_{v\uparrow}(k) \\
    B^+_{v\uparrow}(k) \\
  \end{array}
\right]=\left[
  \begin{array}{c}
    \cos\theta_k \\
    e^{i\frac{kc}{2}}\sin\theta_k \\
  \end{array}
\right], \quad
\left[
  \begin{array}{c}
    A^+_{v\downarrow}(k) \\
    B^+_{v\downarrow}(k) \\
  \end{array}
\right]=\left[
  \begin{array}{c}
    e^{-i\frac{kc}{2}}\sin\theta_k \\
    \cos\theta_k  \\
  \end{array}
\right],
\end{equation}
where $\{\cos(2\theta_k),\sin(2\theta_k)\}=\{\Delta_v/\sqrt{\Delta_v^2+16t^2\cos^2(kl)},
4t/\sqrt{\Delta_v^2+16t^2\cos^2(kl)}\}$.
The solutions for conduction band states can be written as
\begin{align}
|\Phi_{cs}^{k},+\rangle=\frac{1}{\sqrt{M}}\sum_{m=1}^M e^{2ikml}|\Psi^{(2m-1)}_{v},s\rangle, \quad
|\Phi_{cs}^{k},-\rangle=\frac{1}{\sqrt{M}}\sum_{m=1}^M e^{2ikml}|\Psi^{(2m)}_{c},s\rangle.
\end{align}
Their spectrum of energies is
\begin{equation}
E^\pm_{cs}(k)=E_c\pm\sigma_s\frac{\Delta_c}{2}\pm(g_c-2\delta g_c+\sigma_s)\mu_BB-4\sigma_sg_u\mu_BB\cos^2(kl)\pm2\bar{g}_c\mu_BB\cos(2kl).
\end{equation}
A direct calculation demonstrates that $A$-exciton optical transitions are possible only between
$\{|\Phi_{v\uparrow}^{k}\rangle, |\Phi_{c\uparrow}^{k},+\rangle\}$ and $\{|\Phi_{v\downarrow}^{k}\rangle, |\Phi_{c\downarrow}^{k},-\rangle\}$ pairs of states, with the same wave-vector $k$. The corresponding
transitions active in $\sigma^+$ and $\sigma^-$ polarisations respectively and have the same intensities
\begin{equation}
J(k)=|A^+_{v\uparrow}(k)|^2=|B^+_{v\downarrow}(k)|^2=
\frac{1}{2}\Big(1+\frac{\Delta_v}{\sqrt{\Delta_v^2+16t^2\cos^2(kl)}}\Big).
\end{equation}
The $g$-factors of these transitions have opposite signs
\begin{equation}
g_+(k)=-g_-(k)=g_c-2\delta g_c-4g_u\cos^2(kl)+2\bar{g}_c\cos(2kl)-\frac{\Delta_v[g_v-2\delta g_v+2\bar{g}_v\cos(2kl)]}{\sqrt{\Delta_v^2+16t^2\cos^2(kl)}}.
\end{equation}
Next, the averaging of the $g$-factor with corresponding weights gives
\begin{equation}
g^{(\infty)}_{exc}=g_{exc}+4[\delta g_v-\delta g_c-g_u]+\frac{4t^2}{\Delta_v^2}[2g_v+g_u-4\delta g_v -\bar{g}_c+3\bar{g}_v].
\end{equation}

\end{widetext}

\end{document}